\documentclass[fleqn,usenatbib]{mnras}

\usepackage{newtxtext,newtxmath}

\usepackage[T1]{fontenc}

\DeclareRobustCommand{\VAN}[3]{#2}
\let\VANthebibliography\thebibliography
\def\thebibliography{\DeclareRobustCommand{\VAN}[3]{##3}\VANthebibliography}

\usepackage{graphicx}	
\usepackage{amsmath}	
\usepackage{physics}

\title[Detecting FRBs in GW astronomy]{Continuous gravitational wave detection to understand the generation mechanism of fast radio bursts}

\author[S. Kalita and A. Weltman]{
Surajit Kalita\thanks{Corresponding author; E-mail: surajit.kalita@uct.ac.za}
and Amanda Weltman\thanks{E-mail: amanda.weltman@uct.ac.za}
\\
High Energy Physics, Cosmology \& Astrophysics Theory (HEPCAT) Group, Department of Mathematics \& Applied Mathematics, \\University of Cape Town, Cape Town 7700, South Africa
}

\date{Accepted XXX. Received YYY; in original form ZZZ}

\pubyear{2022}

\begin{document}
\label{firstpage}
\pagerange{\pageref{firstpage}--\pageref{lastpage}}
\maketitle

\begin{abstract}
Since the unexpected discovery of fast radio bursts (FRBs), researchers have proposed varied theories and models to explain these phenomena. One such model that has recently been developed incorporates the so-called Gertsenshtein-Zel'dovich (GZ) effect, which states that when gravitational waves traverse a pulsar magnetosphere, a portion of the gravitational radiation is transformed into electromagnetic (EM) radiation. The observed properties of FRBs are consistent with the properties of this EM radiation, implying, remarkably, that the GZ effect can account for both repeating and non-repeating FRBs. If this model is correct, the pulsar's properties should not change over time, and it would continue to emit both EM dipole and gravitational quadrupole radiation for a long period of time. This article targets the gravitational radiation produced by the pulsar mechanism and shows that several proposed gravitational wave detectors can detect these gravitational waves. If such detections are performed in the future from the location of FRBs, it might validate the GZ process for FRB production and potentially rule out several other theories of FRB generation.
\end{abstract}

\begin{keywords}
(transients:) fast radio bursts -- gravitational waves -- stars: magnetic field -- stars: rotation -- (stars:) pulsars: general -- radiation mechanisms: general
\end{keywords}



\section{Introduction}
Fast radio bursts (FRBs) are bright radio transient events~(observed flux is typically on the order of Jy) with approximately millisecond durations. Since their first discovery by~\cite{2007Sci...318..777L}, several radio telescopes, such as the Canadian Hydrogen Intensity Mapping Experiment (CHIME)\footnote{https://chime-experiment.ca/en}, Parkes\footnote{https://www.parkes.atnf.csiro.au/}, Australian Square Kilometre Array Pathfinder (ASKAP)\footnote{https://www.atnf.csiro.au/projects/askap/index.html}, Molonglo Observatory Synthesis Telescope (UTMOST)\footnote{https://astronomy.swin.edu.au/research/utmost/}, Pushchino~\citep{2009PhyU...52.1159D}\footnote{http://www.prao.ru/English/index.php}, etc., have so far detected over 500 FRBs, in between 100\,MHz and 8\,GHz frequency range during the last decade. The majority of these have been detected by CHIME in recent years. The relatively large dispersion measures observed indicate that most FRBs have extragalactic origins. A notable exception is FRB\,200428, which is confirmed to originate from a Galactic magnetar SGR\,1935$+$2154~\citep{2020PASP..132c4202B,2020Natur.587...59B,2020Natur.587...54C}. The anticipated rate of observable FRBs in the entire sky is estimated to be around 1000 per day~\citep{2016MNRAS.460L..30C}.

As more FRBs are identified, our understanding of possible plausible FRB progenitors improves. Since their discovery, several models using neutron stars (NSs), black holes (BHs), and white dwarfs (WDs) have been proposed to explain some features of FRBs; see \cite{2019PhR...821....1P} for a comprehensive review of progenitor models. According to~\cite{2014A&A...562A.137F}, a supramassive rotating NS, which may be formed in a NS-NS merger, collapses to a BH, and the magnetic field lines suddenly shatter. It causes a magnetic shock to occur, and the accelerated electrons travelling with the shock dissipate a considerable portion of their energy in the magnetosphere, resulting in FRBs. Other models account for a binary NS merger~\citep{2013PASJ...65L..12T} or a binary WD merger~\citep{2013ApJ...776L..39K}, or a WD-NS merger~\citep{2020IJAA...10...28L}, in which coherent radio emission is generated either from the entire surface or from the polar region of the combined object at the time of the merger, and this radiation is what we see as FRBs. Theories incorporating magnetars, highly magnetized neutron stars, have gained significant traction since the discovery of an FRB associated to a galactic magnetar as mentioned above \cite{2020Natur.587...59B}. Magnetar origin theories involve different physical mechanisms to produce the bursts including curvature radiation mechanism~\citep{2017MNRAS.468.2726K,2019MNRAS.483L..93L}, starquake mechanisms such as the crustal activity of a magnetar~\citep{2018ApJ...852..140W}, synchrotron maser emission from relativistic, magnetized shocks~\citep{2014MNRAS.442L...9L}, giant flares in soft gamma repeaters~\citep{2014ApJ...797...70K}, etc., are also popular because they can explain various features of FRBs.

Notably, some FRBs are observed to repeat, and many appear to be single events. Hence, the progenitor theories that forecast the repeating FRBs appear to be more promising because they can equally predict the apparently non-repeating ones, suggesting that they may repeat after a long time, or that we are yet to observe their repetitions. Of course, given the range of properties for FRBs observed so far, and the variability of the known host environments, it is plausible, even likely, that there are different types of FRBs, with repeaters and non-repeaters falling into two different classes. Nonetheless, a mechanism that could explain both classes would be compelling.

Recently, \cite{2022arXiv220200032K} suggested a novel generation mechanism based on the Gertsenshtein-Zel'dovich (GZ) effect which implies that when gravitational waves (GWs) propagate through the magnetosphere of a pulsar, a part of their energy is transformed into electromagnetic (EM) radiation in radio frequencies, which we observe as FRBs. They showed that this model can simultaneously explain both repeating and non-repeating FRBs. It is worth noting that this process is reversible, meaning that EM radiation can also be transformed to GWs in the presence of a magnetic field. This was earlier proposed by~\cite{1962JETP...41..113G} and later applied in astrophysics by~\cite{1974JETP...38..652Z}. This is now known as the GZ effect. Using this effect, \cite{2001PhRvD..63d4014P} showed the generation of high-frequency GWs in different media. Later, \cite{2005AIPC..746.1264S} provided a simple demonstration of this effect where x-ray light is converted to GWs separately in the presence of static and alternating magnetic fields. Further, \cite{2015PhyS...90g4059K} showed the direct and inverse effects of the GZ mechanism, thereby estimating the strengths of EM and GWs generated. Eventually, several others have demonstrated how to improve the design sensitivity so that we can detect weaker signals at high frequencies~\citep{2018PhRvD..98f4028Z,2021PhRvD.104b3524H}.

Due to the existence of a number of theories for the progenitor mechanism for FRBs, it is premature to single out only one. Even a substantial increase in FRB detections may not be sufficient to constrain progenitor theories, in which case GW astronomy might play a vital role. In this article, we look into the feasibility of using GW detectors to identify the central object that causes the effect, and determine whether the GZ effect is indeed a driver of FRBs. If the central compact object behaves like a pulsar, which means its rotation and magnetic field axes are not aligned, it can generate continuous GWs. If the object is a White Dwarf (WD), it can emit GWs at a frequency lower than $1\rm\,Hz$ and if it is a Neutron Star (NS), the frequency may exceed $1\rm\,Hz$ due to its smaller size. 

Further, the sensitivity of our present ground-based GW detectors has not yet been experimentally proven to be adequate to detect continuous GWs although they might detect such waves in the future. However, different proposed ground-based or space-based detectors, such as Laser Interferometer Space Antenna~(LISA), Big Bang Observer~(BBO), DECi-hertz Interferometer Gravitational Wave Observatory~(DECIGO), advanced Laser Interferometer Gravitational Wave Observatory~(aLIGO), Einstein Telescope~(ET), Cosmic Explorer~(CE), etc.~\citep{2015CQGra..32a5014M,2019Natur.568..469M,2021NatRP...3..344B}, may detect the continuous GW signal from WD and NS pulsars~\citep{1996A&A...312..675B,2007PhRvD..76h2001A,2014ApJ...785..119A,2020ApJ...896...69K,2021MNRAS.508..842K}. When gravitational radiation is converted to EM radiation due to the GZ effect, the pulsar properties including the magnetic field strength, rotation speed and the pulsar angle, remain the same, and they can continuously emit GWs. Once future detectors are operational and detect a GW signal from the site of the observed FRBs, they can immediately tell us that the central object (in this case a pulsar), is still intact, ruling out some models involving BHs or collisions, and emphasising other theories like the GZ effect.

If the GZ effect is responsible for the formation of FRBs, we can extract the central object's characteristic attributes, such as the magnetic field strength, angular velocity and pulsar angle, from the observed properties of the bursts. In this article, we use these parameters to calculate how long it takes a specific GW detector to detect this signal based on its sensitivity curve. The following is an outline of how this article is organized. In Section~\ref{Sec2}, we first briefly discuss the GZ mechanism and then introduce continuous GWs and their detection. In Section~\ref{Sec3}, we discuss the GW strengths due to the pulsar mechanism for the compact objects associated with the FRBs and thereby estimate the required time to detect these GW signals using various GW detectors. We choose a few typical FRBs to see whether the GW detectors can detect the central object within 1\,yr of their respective operation periods. Finally, we put our concluding remarks in Section~\ref{Sec4} by discussing various results.

\section{Gertsenshtein-Zel'dovich effect and GW detection technique}\label{Sec2}

According to the GZ effect, if GWs move through a transverse magnetic field, an induced EM field is produced; a part of the incident gravitational radiation is converted to EM radiation. Let us consider a pulsar rotating at a frequency $\Omega_\mathrm{rot}$. Note that, by pulsar, we mean that it can either be a WD pulsar or a NS pulsar. Hence, the effective magnetic field at any point in the pulsar magnetosphere at a time $t$ can be written as $\vec{B}(t) = \vec{B}^{(0)} + \delta \vec{B} \sin(\Omega_\mathrm{rot} t)$~\citep{2022arXiv220200032K}. Now, if a GW with frequency $\Omega_g$ and wave number $k_g$ travels in $z$-direction, the two modes of polarization for this GW can be written as
\begin{align}
    h_+ = A_+ e^{i\left(k_g z - \Omega_g t\right)} \quad \text{and} \quad
    h_\times = i A_\times e^{i\left(k_g z - \Omega_g t\right)}.
\end{align}
\begin{figure}
	\centering
	\includegraphics[scale=0.48]{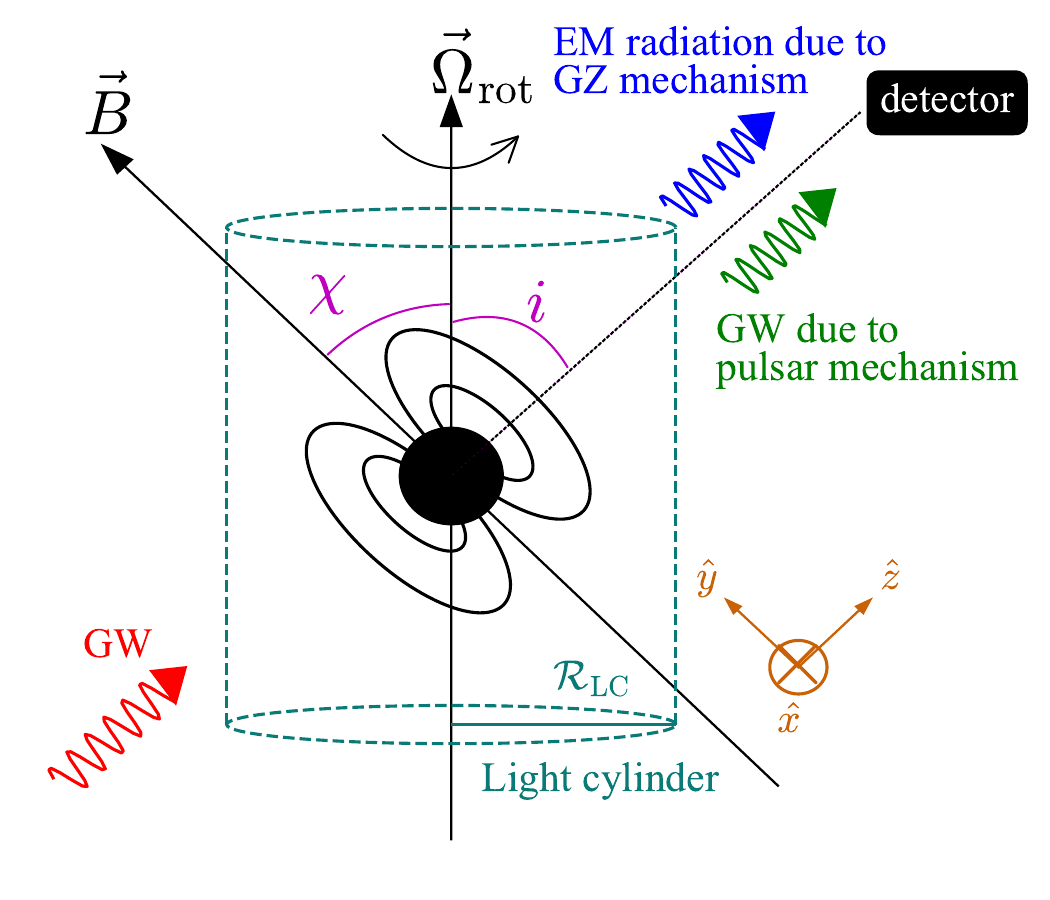}
	\caption{Schematic diagram of a pulsar where magnetic field axis makes an angle of $\chi$ with the rotation axes. The angle between the rotation axis and the detector's line of sight is $i$.}
	\label{Fig: pulsar}
\end{figure}
Let us consider this gravitational radiation falls in the pulsar magnetosphere where the magnetic field is in $y$-direction, i.e. $\vec{B}(t) = \left(0, B^{(0)}_y + \delta B_y \sin(\Omega_\mathrm{rot} t), 0\right)$. It is schematically shown in Figure~\ref{Fig: pulsar}. Now, the background is curved due to the presence of GWs and we can no longer consider a flat Minkowski background. As a result, due to the coupling between the GWs and EM field, the EM field tensor is modified and the resulting electric and magnetic field components are given by the following wave equations~\citep{2022arXiv220200032K}
\begin{align}
    \frac{1}{c^2}\pdv[2]{\tilde{E}_x}{t} -\partial_z^2{\tilde{E}_x} &= f_E(z,t), \\
    \frac{1}{c^2}\pdv[2]{\tilde{B}_y}{t} -\partial_z^2{\tilde{B}_y} &= f_B(z,t),
\end{align}
where
\begin{align*}
    f_E(z,t) &=
    -\frac{A_+B_y^{(0)}k_g\Omega_g}{c}e^{i\left(k_g z - \Omega_g t\right)} \nonumber\\ 
    &- \frac{iA_+ \delta{B_y}k_g}{2c}\left[\Omega_+ e^{i\left(k_g z - \Omega_+ t\right)} - \Omega_- e^{i\left(k_g z - \Omega_- t\right)}\right] \nonumber\\ 
    &- \frac{zA_+ \delta{B_y}\Omega_\mathrm{rot}}{2c^3}\left[\Omega_+^2 e^{i\left(k_g z - \Omega_+ t\right)} + \Omega_-^2 e^{i\left(k_g z - \Omega_- t\right)}\right], \\
    f_B(z,t) &=
    -A_+B_y^{(0)}k_g^2 e^{i\left(k_g z - \Omega_g t\right)} \nonumber \\
    &-\frac{iA_+\delta{B_y}k_g^2\Omega_g}{2}\left[e^{i\left(k_g z - \Omega_+ t\right)} - e^{i\left(k_g z - \Omega_- t\right)}\right] \nonumber\\ 
    &- \frac{iA_+ \delta{B_y}\Omega_\mathrm{rot}}{2c^2}\left[\Omega_+ e^{i\left(k_g z - \Omega_+ t\right)} - \Omega_- e^{i\left(k_g z - \Omega_- t\right)}\right] \nonumber\\ 
    &+ \frac{zA_+ \delta{B_y}\Omega_\mathrm{rot} k_g}{2c^2}\left[\Omega_+^2 e^{i\left(k_g z - \Omega_+ t\right)} - \Omega_-^2 e^{i\left(k_g z - \Omega_- t\right)}\right],
\end{align*}
with $\Omega_{\pm} = \Omega_g \pm \Omega_\mathrm{rot}$. For simplicity, we assume that $A_+=A_\times$. Hence, the resulting EM wave consists of three frequencies: $\Omega_g$ and $\Omega_{\pm}$. Now, for the infalling GWs, if $\Omega_g\gg\Omega_\mathrm{rot}$ such that $\Omega_\pm \approx \Omega_g$, the solutions of the above wave equations are given by
\begin{align}
    \tilde{E}_x &\approx -\frac{1}{2}\left(B_y^{(0)}A_+ - \delta{B_y}A_+ \Omega_\mathrm{rot} t\right)e^{i\left(k_g z - \Omega_g t\right)}, \\
    \tilde{B}_y &\approx -\frac{1}{4}\left(B_y^{(0)}A_+ + 2\delta{B_y}A_+ \Omega_g t\right)e^{i\left(k_g z - \Omega_g t\right)}.
\end{align}
Now, the energy density carried by these induced EM waves is given by
\begin{align}
    \rho_{\mathrm{EM}} = \frac{\abs{\tilde{E}_x}^2 + \abs{\tilde{B}_y}^2 }{8 \pi} \approx \frac{ \abs{A_+}^2 \abs{B^{(0)}_y}^2 }{128 \pi}  \left(5 + 4\xi^2 \Omega_g^2 t^2 + 4 \xi \Omega_g t\right),
\end{align}
with $\xi = \delta{B_y}/B_y^{(0)}$. In our calculations, we use $\xi=0.01$ which is well within the bound given by~\cite{2012A&A...547A...9P}. Similarly, the energy density for the GWs assuming $A_+=A_\times$, is given by
\begin{align}
\rho_{\mathrm{GW}} = \frac{c^2 \Omega_g^2}{32\pi G} \left(\abs{A_+}^2 + \abs{A_\times}^2 \right) = \frac{c^2 \Omega_g^2}{16 \pi G}  \abs{A_+}^2.
\end{align}
Therefore the amount of GW energy converted in EM waves at a point is given by
\begin{align}
    \alpha = \frac{\rho_{\mathrm{EM}}}{\rho_{\mathrm{GW}}} = \frac{5 G \abs{B^{(0)}_y}^2}{8 c^2} \left[\frac{4}{5} \xi^2  \left(\frac{z}{c}\right)^2 + \frac{4}{5}\frac{\xi}{\Omega_g}\frac{z}{c} + \frac{1}{\Omega_g^2}\right],
\end{align}
and the total amount of energy converted from GWs to EM waves due to the entire pulsar magnetosphere is given by
\begin{align}
    \alpha_\mathrm{tot} &= \frac{1}{\mathcal{R}_\mathrm{LC}}\int_{\mathcal{R}_\mathrm{CO}}^{\mathcal{R}_\mathrm{LC}} \alpha \dd{z}\dd{\Omega}\\
    &\approx \frac{5\pi G \abs{B^{(0)}_y}^2}{2c^2}\left[\frac{4}{15}\xi^2\left(\frac{\mathcal{R}_\mathrm{LC}}{c}\right)^2 + \frac{2\xi}{5\Omega_g}\left(\frac{\mathcal{R}_\mathrm{LC}}{c}\right) + \frac{1}{\Omega_g^2}\right],
\end{align}
where $\Omega$ is the solid angle, ${\mathcal{R}_\mathrm{CO}}$ is the radius of the compact object, and $\mathcal{R}_\mathrm{LC}$ is the radius of the pulsar magnetosphere. In general, ${\mathcal{R}_\mathrm{LC}}\gg{\mathcal{R}_\mathrm{CO}}$ and the above integration is computed under this assumption. Moreover, the Poynting vector (peak flux) is given by
\begin{align}
    S_z &= \frac{c}{8\pi}\abs{\Vec{E}\times\Vec{B}} \\
    &= \frac{A_+^2 \abs{B^{(0)}_y}^2 c}{128\pi}\left[\sqrt{\frac{24 c^2 \Omega_g^2 \alpha_\mathrm{tot}}{\pi G \abs{B^{(0)}_y}^2}-51} - \frac{6 c^2 \Omega_g \Omega_\mathrm{rot} \alpha_\mathrm{tot}}{\pi G \abs{B^{(0)}_y}^2}-1\right].\label{Eq: peak flux}
\end{align}

If the infall GWs were generated in the early universe, they might have a wide frequency range. Hence the linear frequency $\nu_g = \Omega_g/2\pi \approx 10^6-10^9\rm\,Hz$ is achievable in such a scenario. Light primordial black holes evaporating before nucleosynthesis, mergers of primordial black holes, capture in primordial black hole haloes, axion annihilation to photons or gravitons, reheating, oscillon production in the early universe, plasma instabilities, exotic compact object binaries, brane-confined matter, etc. are all examples of mechanism that could generate GWs in this frequency range ~\citep{2003PhRvD..68d4017S,2009PhRvL.103k1303A,2013PhRvL.110g1105A,2015PhRvD..92l3009H,2015IJMPD..2430031K,2016JCAP...10..001G,2019EPJC...79.1032E,2021LRR....24....4A,2021JCAP...01..001G,2021JPhCS2081a2009P,2021PhRvD.104j3009S}. On the other hand, for a NS pulsar, the linear frequency $\nu_\mathrm{rot} = \Omega_\mathrm{rot}/2\pi \lesssim 1\rm\,kHz$ and for a WD pulsar, $\nu_\mathrm{rot}\lesssim 1\rm\,Hz$. Therefore, it readily follows the condition $\Omega_g\gg\Omega_\mathrm{rot}$, and the above calculations are valid.

Once the GWs generated instantaneously in the early universe interact and pass through the magnetosphere, EM radiation is produced with a frequency nearly equal to $\Omega_g$. As a result, the radio detectors detect a sudden flash of radiation from the position of the pulsar. Note that according to the GZ mechanism, gravitational radiation gets converted to EM radiation only if the infall waves are perpendicular to the magnetic fields. Thus, even if the infall GWs is quasi-continuous in nature, unless the pulsar position is such that its magnetic axis is perpendicular to the infall GWs, the GZ mechanism will not work. As the pulsar is rotating in a different direction to the magnetic field, only when they become mutually perpendicular to each other, the pulsar magnetosphere can convert GWs to EM radiation. If the detector's line of sight aligns with the direction of infall waves, we see this EM radiation as a flash of light and thus it can explain the origin of repeating FRBs. 

In the next section, we show in a couple of examples that this radiation has a flux and pulse width similar to those of observed FRBs; therefore GZ effect can explain the origin of these FRBs. The pulse width equals the time needed for the radiation to cross the entire magnetosphere. Once the GWs pass the magnetosphere, the pulsar continues, emitting both EM and GW radiations for a long time. The pulsar can be detected by EM telescopes if it is near enough. However, if it is far away, then it is a challenge for EM telescopes to detect these pulsars. Notably, it has been reported before \citep{2007PhRvD..76h2001A,2014ApJ...785..119A}, that different proposed GW detectors will be able to detect continuous GWs from distant pulsars in the future.

The two GW polarization modes emitted from a pulsar are given by~\citep{Maggiore}
\begin{align}
	\tilde{h}_+ &= \tilde{A}_{+,1}\cos\left({\Omega_\text{rot} t}\right) + \tilde{A}_{+,2}\cos\left({2\Omega_\text{rot} t}\right), \\
	\tilde{h}_\times &= \tilde{A}_{\times,1}\sin\left({\Omega_\text{rot} t}\right) + \tilde{A}_{\times,2}\sin\left({2\Omega_\text{rot} t}\right),
\end{align}
where
\begin{equation}
    \begin{aligned}
    	\tilde{A}_{+,1} &= \tilde{h}_0 \sin 2\chi \sin i \cos i, \\
    	\tilde{A}_{+,2} &= 2\tilde{h}_0 \sin^2\chi (1 + \cos^2 i), \\
    	\tilde{A}_{\times,1} &= \tilde{h}_0 \sin 2\chi \sin i, \\
    	\tilde{A}_{\times,2} &= 4\tilde{h}_0 \sin^2\chi \cos i,
\end{aligned}
\end{equation}
with
\begin{equation}
	\tilde{h}_0 = \frac{G}{c^4}\frac{\Omega_\text{rot}^2 \epsilon I_2}{d}.
\end{equation}
Here $I_2$ represents the moment of inertia of the object about the magnetic field axis and $I_3$ represents the same with respect to the axis perpendicular to the magnetic field axis, such that ellipticity is defined as $\epsilon = \abs{I_2-I_3}/I_2$. The magnetic field axis and the detector's line of sight create an angle with the rotation axis of $\chi$ and $i$ respectively. $d$ is the distance between the GW detector and the pulsar. Note that we distinguish between the infall GWs produced in the early universe and the GWs produced by a compact object by using a  `tilde' for the latter. Since pulsars emit both EM and gravitational radiations, they are associated with the EM dipole and gravitational quadrupole luminosities, which are respectively given by~\citep{1979PhRvD..20..351Z,2000MNRAS.313..217M,2006ApJ...648L..51S,2015ApJ...801L..19P}
\begin{align}
    L_\text{D} &= \frac{2B_\mathrm{p}^2 R_\mathrm{p}^6 \Omega_\mathrm{rot}^4}{3c^3} \left(1+\sin^2\chi\right),\\
    L_\text{GW} &= \frac{2G}{5c^5} \epsilon^2 I_2^2 \Omega_\mathrm{rot}^6 \sin^2\chi \left(1+15\sin^2\chi\right),
\end{align}
with $R_\mathrm{p}$ being the stellar radius at the pole where the magnetic field strength is $B_\mathrm{p}$. As a result, $\Omega_\mathrm{rot}$ and $\chi$ decrease over time and their variations are given by~\citep{2000MNRAS.313..217M}
\begin{align}\label{Eq: radiation1}
    I_\mathrm{rot}\dv{\Omega_\mathrm{rot}}{t} &= -\frac{2G}{5c^5} \epsilon^2 I_2^2 \Omega_\mathrm{rot}^5 \sin^2\chi \left(1+15\sin^2\chi\right) \nonumber \\
    &\quad - \frac{2B_\mathrm{p}^2 R_\mathrm{p}^6 \Omega_\mathrm{rot}^3}{3c^3}\left(1+\sin^2\chi\right),\\
    \label{Eq: radiation2}
    I_\mathrm{rot} \dv{\chi}{t} &= -\frac{12G}{5c^5} \epsilon^2 I_2^2 \Omega_\mathrm{rot}^4 \sin^3\chi \cos\chi - \frac{B_\mathrm{p}^2 R_\mathrm{p}^6 \Omega_\mathrm{rot}^2}{3c^3}\sin{2\chi},
\end{align}
where $I_\mathrm{rot}$ is the moment of inertia of the compact object about the rotation axis. Because the GWs are emitted at two frequencies, when a GW detector detects such a signal, the signal-to-noise-ratio ($\mathrm{S/N}$) is given by~\citep{Maggiore}
\begin{equation}\label{Eq: SNR}
    \mathrm{S/N} = \sqrt{\mathrm{S/N}_{\Omega}^2 + \mathrm{S/N}_{2\Omega}^2} ,
\end{equation}
where
\begin{align}\label{Eq: SNR v11}
    \langle \mathrm{S/N}_{\Omega}^2\rangle &= \frac{\sin^2{\zeta}}{100}\frac{h_0^2 T \sin^2{2\chi(t)}}{S_\mathrm{n}(\nu_\mathrm{rot}(t))},\\
    \label{Eq: SNR v12}
    \langle \mathrm{S/N}_{2\Omega}^2\rangle &= \frac{4\sin^2{\zeta}}{25}\frac{h_0^2 T \sin^4{\chi(t)}}{S_\mathrm{n}(2\nu_\mathrm{rot}(t))}.
\end{align}
Here the angle between the interferometer arms is $\zeta$, and the detector's power spectral density (PSD) at the frequency $\nu_\mathrm{rot}$ is $S_\mathrm{n}(\nu_\mathrm{rot})$. The PSD data for several detectors are collected from~\cite{2015CQGra..32a5014M} and ~\cite{2020PhRvD.102f3021H}\footnote{\url{http://gwplotter.com}}. Given the proposed equilateral triangular design of LISA, we assume $\zeta = 60^\circ$ in our calculations while considering LISA and $\zeta=90^\circ$ for LIGO detectors. One can in principle use a time-stacking approach in which the whole observation time is divided into a number of time-stacks. In comparison to a long-term coherent search, an incoherent search employing a time-stacking technique is computationally efficient~\citep{2000PhRvD..61h2001B,2005PhRvD..72d2004C}. As a result, this stacking method can be used to search the entire sky for unknown pulsars~\citep{2012JPhCS.354a2010L}. However, because for most FRBs' angular positions are known, in this work, we do not consider this technique. Furthermore, $\langle\mathrm{S/N}\rangle\gtrsim 5$ is necessary to detect a continuous GW signal for a localised source with more than $95\%$ detection efficiency~\citep{2011MNRAS.415.1849P}.

\section{Detection of GW signal from compact objects producing FRBs}\label{Sec3}

In this section, we consider a few typical FRBs from the CHIME\footnote{\url{https://www.chime-frb.ca/catalog}} and FRBCAT\footnote{\url{https://www.frbcat.org}} catalogues~\citep{2021ApJS..257...59C,2016PASA...33...45P}. The physical parameters of the compact object, such as its rotation rate and magnetic field strength, are obtained by combining FRB's observed attributes along with the GZ effect. Later, we use these quantities to estimate the time required for different GW detectors to detect the continuous GW signal emitted by the pulsar.

We choose specific FRBs from the catalogue to exemplify our analysis. 

\subsection{FRB\,160920}
This FRB was observed by the Pushchino Radio Astronomy Observatory. It has the highest pulse width in the catalogue to date, which will give us the lowest $\Omega_\mathrm{rot}$ and thus put us in the frequency range for LISA. It was observed at $111\rm\,MHz$ frequency with a pulse width $\delta=5\rm\,s$ and peak flux $=0.22\rm\,Jy$. According to the GZ model, the pulse width is the time required for GWs to cross the entire pulsar magnetosphere. Therefore, the radius of the light cylinder is given by
\begin{equation}
    \mathcal{R}_\mathrm{LC} = \frac{\delta c}{2} = 7.49\times10^{10}\rm\,cm.
\end{equation}
Hence the angular speed of the compact object is given by
\begin{equation}
    \Omega_\mathrm{rot} = \frac{c}{\mathcal{R}_\mathrm{LC}} = \frac{2}{\delta} = 0.4\rm\,rad\,s^{-1},
\end{equation}
and thus the linear frequency is
\begin{equation}
    \nu_\mathrm{rot} = \frac{\Omega_\mathrm{rot}}{2\pi} = 0.064\rm\,Hz.
\end{equation}
\begin{figure}
	\centering
	\includegraphics[scale=0.48]{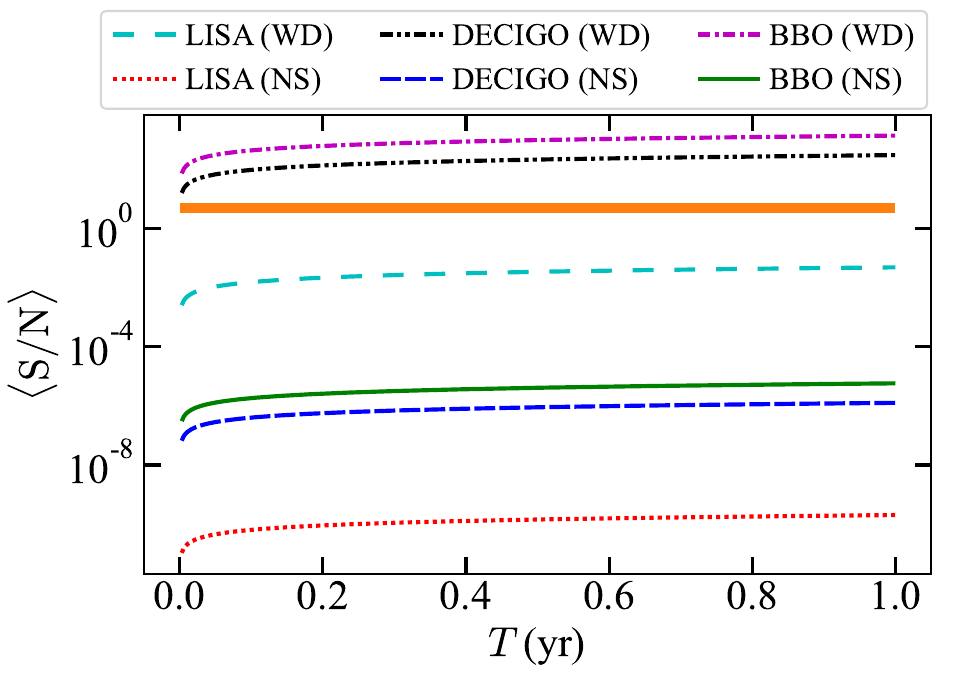}
	\caption{$\mathrm{S/N}$ as a function of integration time for FRB\,160920 assuming $\chi(t=0)=45^\circ$. The thick orange line corresponds to $\langle\mathrm{S/N}\rangle\approx5$.}
	\label{Fig: FRB160920}
\end{figure}
\begin{figure}
	\centering
	\includegraphics[scale=0.48]{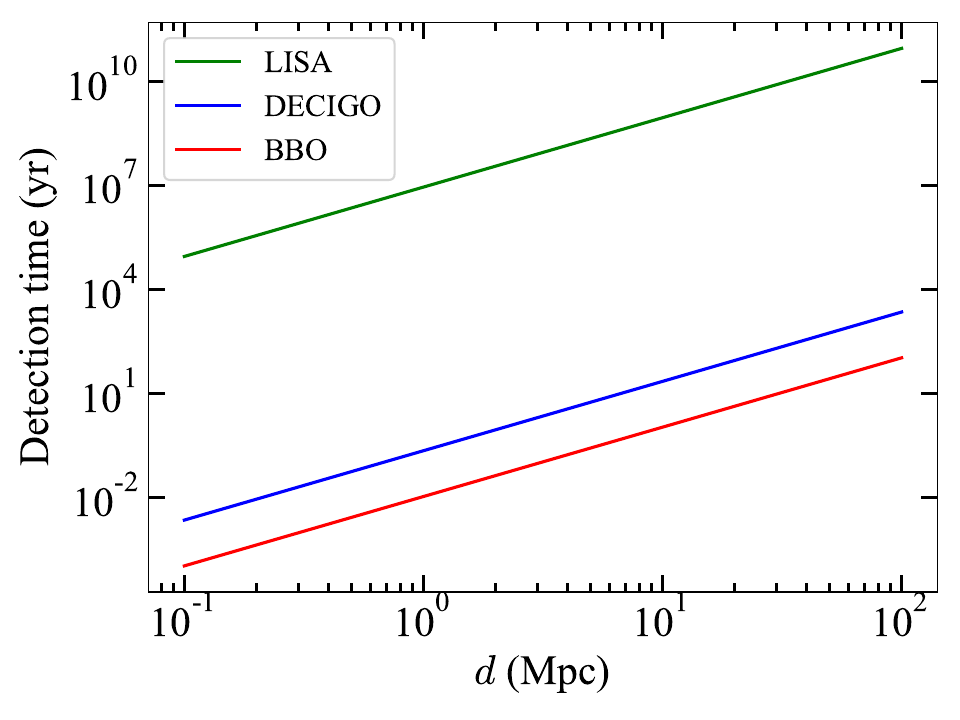}
	\caption{Detection time as a function of the distance to FRB\,160920 assuming it to be a WD for a detection threshold $\mathrm{S/N}=5$.}
	\label{Fig: FRB160920_new}
\end{figure}

Now, because it is observed at $111\rm\,MHz$ frequency with peak flux $=0.22\rm\,Jy$, we have $\nu_g=111\rm\,MHz$ and $S_z/\nu_g=0.22\rm\,Jy$. From Equation~\eqref{Eq: peak flux}, it is evident that only unknown quantities are $\abs{B^{(0)}_y}$ and $A_+$. Assuming $A_+=10^{-24}$, it turns out that $\abs{B^{(0)}_y} = 5.5\times10^8\rm\,G$. Note that such GW strain can be produced by various cosmological mechanisms mentioned in the previous section. For a detailed discussion on the strength of GWs by these phenomena, one may look at the reviews by~\cite{2015IJMPD..2430031K} and~\cite{2021LRR....24....4A}. It is worth noting that the rotation frequency for this particular case can be attained both by a WD and a NS. Similarly, a WD or a NS can also achieve this desired surface magnetic field value. Using these magnetic fields and rotation parameters, we model the WDs and NSs using the {\sc xns} code (version 3.0)\footnote{\url{http://www.arcetri.astro.it/science/ahead/XNS/code.html}} to obtain their $\epsilon$ for the given magnetic field value. A brief discussion on the {\sc xns} code configuration is provided in appendix~\ref{appendix A}. Moreover, the rotation frequency of this object turns out to be less than 1\,Hz. As a result, the currently operational ground-based GW detectors are ineffective and we require futuristic space-based detectors, such as LISA, DECIGO, BBO, and others. 

We now use the PSD for some of these GW detectors and estimate the required time to observe this compact object by the respective detectors. Figure~\ref{Fig: FRB160920} depicts the cumulative $\mathrm{S/N}$ over 1\,yr observation period when the object is 100\,kpc away. It turns out that if the source is a NS, no proposed detector can detect it. However, if it is a WD, some detectors, such as BBO and DECIGO, can detect it within 1\,yr of the observation period. Moreover, because $\mathrm{S/N}\propto 1/d$, from Figure~\ref{Fig: FRB160920_new}, it is also evident that if the source is a WD and it is extra-galactic with the distance being $\mathcal{O}\rm(Mpc)$, still both BBO and DECIGO may detect it within 1\,yr of the observation period.

\subsection{FRB\,180817A}
\begin{figure}
	\centering
	\includegraphics[scale=0.48]{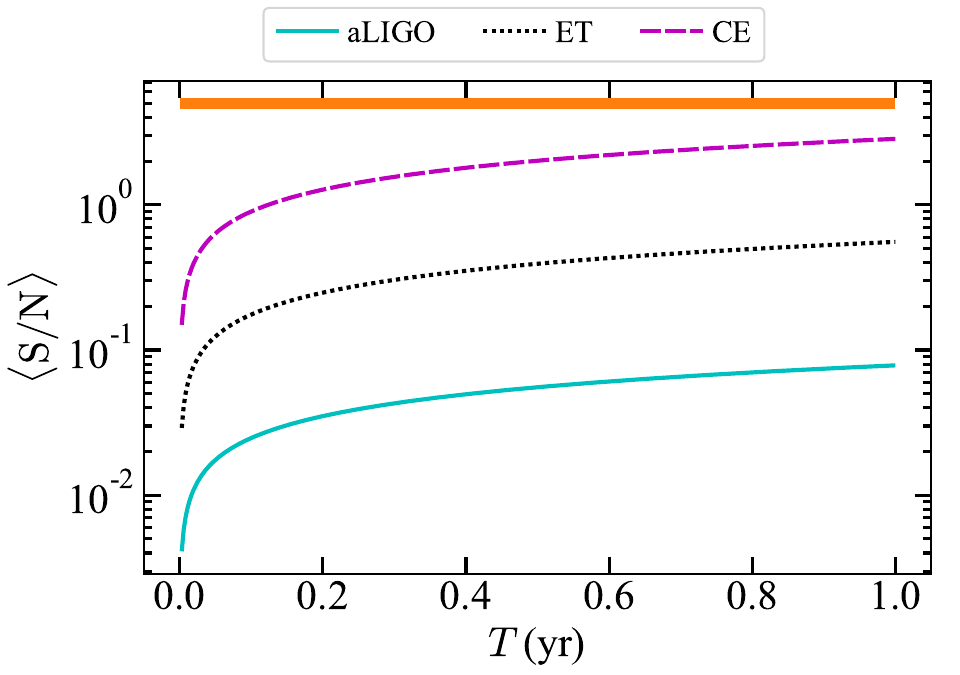}
	\caption{$\mathrm{S/N}$ as a function of integration time for FRB\,180817A.}
	\label{Fig: FRB180817}
\end{figure}
This FRB was observed by the CHIME telescope. Its observed $\delta=0.01769\rm\,s$, $S_z/\nu_g=2.4\rm\,Jy$, and $\nu_g=501.1\rm\,MHz$. Using similar calculations to the aforementioned FRB, it turns out that $\mathcal{R}_\mathrm{LC} = 2.65\times10^8\rm\,cm$, $\Omega_\mathrm{rot} = 113.1\rm\,rad\,s^{-1}$, and $\nu_\mathrm{rot} = 18.0\rm\,Hz$. This is the reason we choose this FRB as its rotation frequency lies approximately at the most sensitive portion of the CE and ET detectors. Because this object cannot be a WD due to its high rotation, we perform all of the necessary calculations for a NS. To match the observed flux, if $A_+ = 10^{-24}$, it turns out that $\abs{B^{(0)}_y} = 3.0\times10^{10}\rm\,G$. 

Substituting these numbers in the code with $d=100$\,kpc assuming the toroidal field component is stronger than the poloidal one at the core, we calculate the $\mathrm{S/N}$ for some detectors as shown in Figure~\ref{Fig: FRB180817}. It is evident that no detector can detect this signal within 1\,yr of observation period. Furthermore, if the incoming GWs on the pulsar have a weak strength, say $A_+=10^{-29}$, we require $\abs{B^{(0)}_y} = 3.0\times10^{15}\rm\,G$ to match the observed flux, implying that it is a magnetar in this scenario. Since the magnetic field is strong, it increases the deformation of the compact object, and hence $\epsilon$ increases, resulting in higher $\mathrm{S/N}$.
\begin{figure}
	\centering
	\includegraphics[scale=0.48]{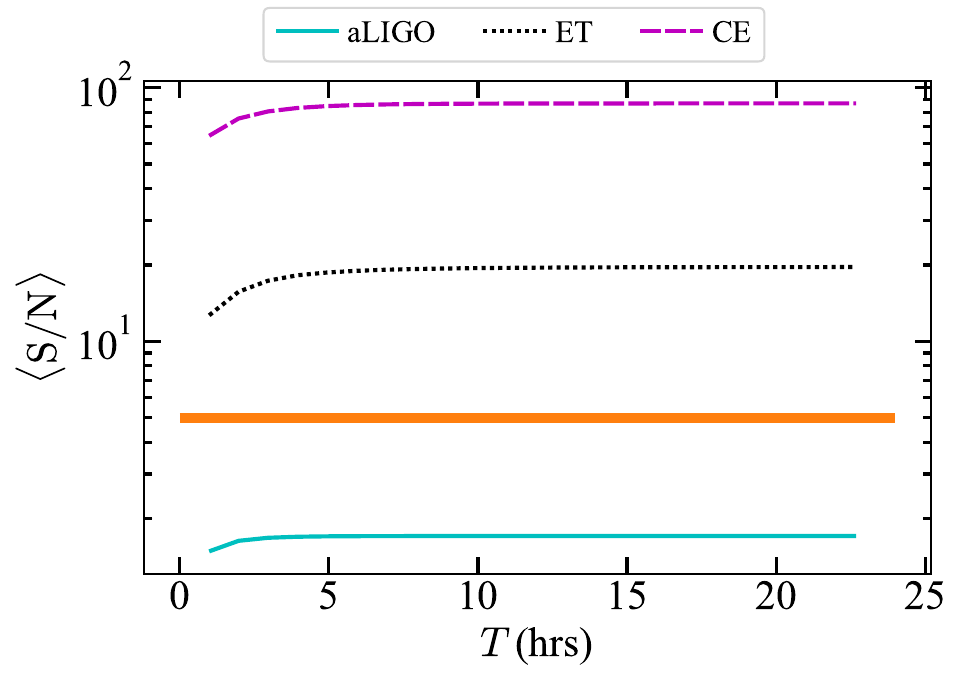}
	\caption{Same as Figure~\ref{Fig: FRB180817} except that now the surface field of the pulsar is $3.0\times10^{15}\rm\,G$.}
	\label{Fig: FRB180817_2}
\end{figure}
Because of the huge surface field, it has a large $L_\text{D}$, and hence its spin-down rate is very high. As a result, both $\chi$ and $\Omega_\mathrm{rot}$ decrease very quickly. Figure~\ref{Fig: FRB180817_2} shows the $\mathrm{S/N}$ for such a magnetar. Because $\chi$ and $\Omega_\mathrm{rot}$ decrease rapidly, the amplitudes of $\tilde{h}_+$ and $\tilde{h}_\times$ similarly fall very fast. Hence, in this scenario, $\mathrm{S/N}$ first increases and then remains nearly constant. It is also evident that the CE and ET detectors may detect the signal almost instantly while aLIGO would still be unable to detect this signal.

\section{Discussion and conclusions}\label{Sec4}

Using the GZ effect is relatively novel for pulsar astronomy. In this paper, we have considered two typical FRBs. From their measured pulse widths, we have calculated the rotation period of the pulsar (be it a white dwarf or a neutron star), whose magnetosphere is responsible for the GZ effect. Further, from their peak flux and the frequency at which the FRB is observed, we have calculated the surface magnetic field of the pulsar. Note that, apart from all of the observed parameters, the only unknown quantity is $A_+$, which is inversely proportional to $\abs{B^{(0)}_y}$. We have chosen $A_+$ in such a way that it can account for ordinary (less magnetic) WDs and NSs as well as the highly magnetized ones, like magnetars. We have used this magnetic field value to execute the {\sc xns} code in order to determine the structure of the pulsar, and consequently, the $\mathrm{S/N}$ of the GW signal, assuming the initial $\chi$ to be $45^\circ$ as illustrated in Figure~\ref{Fig: FRB160920}--\ref{Fig: FRB180817_2}. There might be additional observational techniques, such as the Hough transform, resampling methods, etc.~\citep{2010PhRvD..81h4032P,2011BASI...39..181D}, which might be computationally favourable, but discussion about them is beyond the scope of this paper. We only illustrate here that it is possible to shed light on the nature of FRBs using gravitational wave observations. 

We have shown that the LIGO and LISA cannot detect continuous GWs from the isolated WDs or NSs bearing the configurations (rotation frequencies, magnetic fields, and distances) suggested in the aforementioned examples. However, if the rotation frequency of the source is such that it falls in the LISA-frequency range, only BBO and DECIGO would detect the gravitational radiation within 1\,yr of the observation period, provided the source is a WD even though it is extra-galactic. On the other hand, if the source is a NS and rotates faster, such that its rotation frequency falls in the LIGO-frequency range, then CE and ET detectors can detect the gravitational radiation depending on the surface magnetic field. They can detect the GWs if the source is a magnetar with the surface field around $3\times10^{15}\rm\,G$.

In this paper, we have outlined a potential scenario for utilizing GW astronomy to confirm or refute the GZ effect as a progenitor for FRBs. We have selected a few typical FRBs and computed the GW signal strengths based on their various observed features, assuming that the GZ effect is solely responsible for the formation of FRBs. It is worth noting that this theory differs from other FRB theories, like mergers and other related theories. In the case of a merger, the generated GW signal is instantaneous. Thus, if those theories are responsible for the detected FRBs, we can no longer detect the GW signal generated at the time of the merger. However, according to the GZ effect, an infalling GW radiation can be converted to EM radiation due to the pulsar magnetosphere, and we observe it as FRBs. Thus the pulsar continues to rotate as it does since its birth and is capable of continuously emitting gravitational radiation. This is a distinct signature. In this paper, our target is to detect this GW radiation. In the future, if the proposed GW detectors detect any continuous GW signal from the site of FRBs, this will immediately imply that the merger-like theories cannot explain all FRBs and thus provide significant support for the GZ theory.

\appendix
\section{Brief discussion on the {\sc xns} code configuration}\label{appendix A}
{\sc xns} code was developed based on the algorithm that solves the time-independent general relativistic magnetohydrodynamic (GRMHD) equations to establish magneto-hydrostatic equilibrium for the compact object~\citep{2014MNRAS.439.3541P,2017MNRAS.470.2469P}. It determines the equilibrium structure of uniformly or differentially rotating compact objects together with purely toroidal or poloidal magnetic fields. This code was originally developed to understand the structure of NSs, but we changed it appropriately to handle WD configurations also. Detailed discussions on configuring NSs through {\sc xns} code are given by ~\cite{2014MNRAS.439.3541P,2017MNRAS.470.2469P} and those for WDs are given by~\cite{2015JCAP...05..016D,2019MNRAS.490.2692K,2020ApJ...896...69K,2021MNRAS.508..842K}. 

In this work, because we know the surface magnetic fields, we first run the code assuming a purely poloidal configuration. We find that  the central poloidal field is nearly 2 orders of magnitude larger than the surface field. Further,~\cite{2014MNRAS.437..675W} showed that if the compact object was born with a dominant $\Omega$-dynamo action, its toroidal field component would eventually be nearly 2 orders of magnitude larger than the poloidal field. Thus for our case, the central toroidal field could be as large as $5.5\times10^{12}\rm\,G$ when the surface field is approximately $5.5\times10^{8}\rm\,G$. Note that the central field is primarily responsible for the change in the shape of the compact object and thus it is the determining factor for $\epsilon$. Hence, we further run our code with this particular central toroidal field component to obtain $\epsilon$. The surface poloidal field component only contributes to the dipole luminosity. Although it is known that a star would be unstable for purely toroidal or poloidal field configurations, we need to make this adjustment as the code cannot simultaneously handle rotation and a suitably mixed field configuration. Note that this field value and rotation speed are such that they are well within the bound proposed by~\cite{1989MNRAS.237..355K} and~\cite{2009MNRAS.397..763B}; thus making the object to be in a stable equilibrium condition.

\section*{Acknowledgements}

We thank the anonymous reviewer for their constructive comments to improve the quality of this manuscript. We further thank Marisa Geyer for her useful comments and suggestions to improve the quality of the paper. We gratefully acknowledge support from the University of Cape Town Vice Chancellor’s Future Leaders 2030 Awards programme which has generously funded this research and support from the South African Research Chairs Initiative of the Department of Science and Technology and the National Research Foundation.

\section*{Data availability}
The data underlying this paper will be shared on a reasonable request to the corresponding author.



\bibliographystyle{mnras}
\bibliography{bibliography}

\begin{thebibliography}{}
\makeatletter
\relax
\def\mn@urlcharsother{\let\do\@makeother \do\$\do\&\do\#\do\^\do\_\do\%\do\~}
\def\mn@doi{\begingroup\mn@urlcharsother \@ifnextchar [ {\mn@doi@}
  {\mn@doi@[]}}
\def\mn@doi@[#1]#2{\def\@tempa{#1}\ifx\@tempa\@empty \href
  {http://dx.doi.org/#2} {doi:#2}\else \href {http://dx.doi.org/#2} {#1}\fi
  \endgroup}
\def\mn@eprint#1#2{\mn@eprint@#1:#2::\@nil}
\def\mn@eprint@arXiv#1{\href {http://arxiv.org/abs/#1} {{\tt arXiv:#1}}}
\def\mn@eprint@dblp#1{\href {http://dblp.uni-trier.de/rec/bibtex/#1.xml}
  {dblp:#1}}
\def\mn@eprint@#1:#2:#3:#4\@nil{\def\@tempa {#1}\def\@tempb {#2}\def\@tempc
  {#3}\ifx \@tempc \@empty \let \@tempc \@tempb \let \@tempb \@tempa \fi \ifx
  \@tempb \@empty \def\@tempb {arXiv}\fi \@ifundefined
  {mn@eprint@\@tempb}{\@tempb:\@tempc}{\expandafter \expandafter \csname
  mn@eprint@\@tempb\endcsname \expandafter{\@tempc}}}

\bibitem[\protect\citeauthoryear{{Aasi} et~al.,}{{Aasi}
  et~al.}{2014}]{2014ApJ...785..119A}
{Aasi} J.,  et~al., 2014, \mn@doi [\apj] {10.1088/0004-637X/785/2/119}, \href
  {https://ui.adsabs.harvard.edu/abs/2014ApJ...785..119A} {785, 119}

\bibitem[\protect\citeauthoryear{{Abbott} et~al.,}{{Abbott}
  et~al.}{2007}]{2007PhRvD..76h2001A}
{Abbott} B.,  et~al., 2007, \mn@doi [\prd] {10.1103/PhysRevD.76.082001}, \href
  {https://ui.adsabs.harvard.edu/abs/2007PhRvD..76h2001A} {76, 082001}

\bibitem[\protect\citeauthoryear{{Aggarwal} et~al.,}{{Aggarwal}
  et~al.}{2021}]{2021LRR....24....4A}
{Aggarwal} N.,  et~al., 2021, \mn@doi [Living Reviews in Relativity]
  {10.1007/s41114-021-00032-5}, \href
  {https://ui.adsabs.harvard.edu/abs/2021LRR....24....4A} {24, 4}

\bibitem[\protect\citeauthoryear{{Anantua}, {Easther}  \& {Giblin}}{{Anantua}
  et~al.}{2009}]{2009PhRvL.103k1303A}
{Anantua} R.,  {Easther} R.,   {Giblin} John~T. J.,  2009, \mn@doi [\prl]
  {10.1103/PhysRevLett.103.111303}, \href
  {https://ui.adsabs.harvard.edu/abs/2009PhRvL.103k1303A} {103, 111303}

\bibitem[\protect\citeauthoryear{{Arvanitaki} \& {Geraci}}{{Arvanitaki} \&
  {Geraci}}{2013}]{2013PhRvL.110g1105A}
{Arvanitaki} A.,  {Geraci} A.~A.,  2013, \mn@doi [\prl]
  {10.1103/PhysRevLett.110.071105}, \href
  {https://ui.adsabs.harvard.edu/abs/2013PhRvL.110g1105A} {110, 071105}

\bibitem[\protect\citeauthoryear{{Bailes} et~al.,}{{Bailes}
  et~al.}{2021}]{2021NatRP...3..344B}
{Bailes} M.,  et~al., 2021, \mn@doi [Nature Reviews Physics]
  {10.1038/s42254-021-00303-8}, \href
  {https://ui.adsabs.harvard.edu/abs/2021NatRP...3..344B} {3, 344}

\bibitem[\protect\citeauthoryear{{Bochenek}, {McKenna}, {Belov}, {Kocz},
  {Kulkarni}, {Lamb}, {Ravi}  \& {Woody}}{{Bochenek}
  et~al.}{2020a}]{2020PASP..132c4202B}
{Bochenek} C.~D.,  {McKenna} D.~L.,  {Belov} K.~V.,  {Kocz} J.,  {Kulkarni}
  S.~R.,  {Lamb} J.,  {Ravi} V.,   {Woody} D.,  2020a, \mn@doi [\pasp]
  {10.1088/1538-3873/ab63b3}, \href
  {https://ui.adsabs.harvard.edu/abs/2020PASP..132c4202B} {132, 034202}

\bibitem[\protect\citeauthoryear{{Bochenek}, {Ravi}, {Belov}, {Hallinan},
  {Kocz}, {Kulkarni}  \& {McKenna}}{{Bochenek}
  et~al.}{2020b}]{2020Natur.587...59B}
{Bochenek} C.~D.,  {Ravi} V.,  {Belov} K.~V.,  {Hallinan} G.,  {Kocz} J.,
  {Kulkarni} S.~R.,   {McKenna} D.~L.,  2020b, \mn@doi [\nat]
  {10.1038/s41586-020-2872-x}, \href
  {https://ui.adsabs.harvard.edu/abs/2020Natur.587...59B} {587, 59}

\bibitem[\protect\citeauthoryear{{Bonazzola} \& {Gourgoulhon}}{{Bonazzola} \&
  {Gourgoulhon}}{1996}]{1996A&A...312..675B}
{Bonazzola} S.,  {Gourgoulhon} E.,  1996, \aap, \href
  {https://ui.adsabs.harvard.edu/abs/1996A&A...312..675B} {312, 675}

\bibitem[\protect\citeauthoryear{{Brady} \& {Creighton}}{{Brady} \&
  {Creighton}}{2000}]{2000PhRvD..61h2001B}
{Brady} P.~R.,  {Creighton} T.,  2000, \mn@doi [\prd]
  {10.1103/PhysRevD.61.082001}, \href
  {https://ui.adsabs.harvard.edu/abs/2000PhRvD..61h2001B} {61, 082001}

\bibitem[\protect\citeauthoryear{{Braithwaite}}{{Braithwaite}}{2009}]{2009MNRAS.397..763B}
{Braithwaite} J.,  2009, \mn@doi [\mnras] {10.1111/j.1365-2966.2008.14034.x},
  \href {http://adsabs.harvard.edu/abs/2009MNRAS.397..763B} {397, 763}

\bibitem[\protect\citeauthoryear{{CHIME/FRB Collaboration} et~al.,}{{CHIME/FRB
  Collaboration} et~al.}{2020}]{2020Natur.587...54C}
{CHIME/FRB Collaboration} et~al., 2020, \mn@doi [\nat]
  {10.1038/s41586-020-2863-y}, \href
  {https://ui.adsabs.harvard.edu/abs/2020Natur.587...54C} {587, 54}

\bibitem[\protect\citeauthoryear{{CHIME/FRB Collaboration} et~al.,}{{CHIME/FRB
  Collaboration} et~al.}{2021}]{2021ApJS..257...59C}
{CHIME/FRB Collaboration} et~al., 2021, \mn@doi [\apjs]
  {10.3847/1538-4365/ac33ab}, \href
  {https://ui.adsabs.harvard.edu/abs/2021ApJS..257...59C} {257, 59}

\bibitem[\protect\citeauthoryear{{Champion} et~al.,}{{Champion}
  et~al.}{2016}]{2016MNRAS.460L..30C}
{Champion} D.~J.,  et~al., 2016, \mn@doi [\mnras] {10.1093/mnrasl/slw069},
  \href {https://ui.adsabs.harvard.edu/abs/2016MNRAS.460L..30C} {460, L30}

\bibitem[\protect\citeauthoryear{{Cutler}, {Gholami}  \& {Krishnan}}{{Cutler}
  et~al.}{2005}]{2005PhRvD..72d2004C}
{Cutler} C.,  {Gholami} I.,   {Krishnan} B.,  2005, \mn@doi [\prd]
  {10.1103/PhysRevD.72.042004}, \href
  {https://ui.adsabs.harvard.edu/abs/2005PhRvD..72d2004C} {72, 042004}

\bibitem[\protect\citeauthoryear{{Dagkesamanskii}}{{Dagkesamanskii}}{2009}]{2009PhyU...52.1159D}
{Dagkesamanskii} R.~D.,  2009, \mn@doi [Physics Uspekhi]
  {10.3367/UFNe.0179.200911i.1225}, \href
  {https://ui.adsabs.harvard.edu/abs/2009PhyU...52.1159D} {52, 1159}

\bibitem[\protect\citeauthoryear{{Das} \& {Mukhopadhyay}}{{Das} \&
  {Mukhopadhyay}}{2015}]{2015JCAP...05..016D}
{Das} U.,  {Mukhopadhyay} B.,  2015, \mn@doi [\jcap]
  {10.1088/1475-7516/2015/05/016}, \href
  {http://adsabs.harvard.edu/abs/2015JCAP...05..016D} {5, 016}

\bibitem[\protect\citeauthoryear{{Dhurandhar}}{{Dhurandhar}}{2011}]{2011BASI...39..181D}
{Dhurandhar} S.~V.,  2011, Bulletin of the Astronomical Society of India, \href
  {https://ui.adsabs.harvard.edu/abs/2011BASI...39..181D} {39, 181}

\bibitem[\protect\citeauthoryear{{Ejlli}, {Ejlli}, {Cruise}, {Pisano}  \&
  {Grote}}{{Ejlli} et~al.}{2019}]{2019EPJC...79.1032E}
{Ejlli} A.,  {Ejlli} D.,  {Cruise} A.~M.,  {Pisano} G.,   {Grote} H.,  2019,
  \mn@doi [European Physical Journal C] {10.1140/epjc/s10052-019-7542-5}, \href
  {https://ui.adsabs.harvard.edu/abs/2019EPJC...79.1032E} {79, 1032}

\bibitem[\protect\citeauthoryear{{Falcke} \& {Rezzolla}}{{Falcke} \&
  {Rezzolla}}{2014}]{2014A&A...562A.137F}
{Falcke} H.,  {Rezzolla} L.,  2014, \mn@doi [\aap]
  {10.1051/0004-6361/201321996}, \href
  {https://ui.adsabs.harvard.edu/abs/2014A&A...562A.137F} {562, A137}

\bibitem[\protect\citeauthoryear{Gertsenshtein}{Gertsenshtein}{1962}]{1962JETP...41..113G}
Gertsenshtein M.~E.,  1962, Soviet Journal of Experimental and Theoretical
  Physics, 41, 113

\bibitem[\protect\citeauthoryear{{Giudice}, {McCullough}  \&
  {Urbano}}{{Giudice} et~al.}{2016}]{2016JCAP...10..001G}
{Giudice} G.~F.,  {McCullough} M.,   {Urbano} A.,  2016, \mn@doi [\jcap]
  {10.1088/1475-7516/2016/10/001}, \href
  {https://ui.adsabs.harvard.edu/abs/2016JCAP...10..001G} {2016, 001}

\bibitem[\protect\citeauthoryear{{Guo}, {Sinha}, {Vagie}  \& {White}}{{Guo}
  et~al.}{2021}]{2021JCAP...01..001G}
{Guo} H.-K.,  {Sinha} K.,  {Vagie} D.,   {White} G.,  2021, \mn@doi [\jcap]
  {10.1088/1475-7516/2021/01/001}, \href
  {https://ui.adsabs.harvard.edu/abs/2021JCAP...01..001G} {2021, 001}

\bibitem[\protect\citeauthoryear{{Herman}, {F{\.z}zfa}, {Lehoucq}  \&
  {Clesse}}{{Herman} et~al.}{2021}]{2021PhRvD.104b3524H}
{Herman} N.,  {F{\.z}zfa} A.,  {Lehoucq} L.,   {Clesse} S.,  2021, \mn@doi
  [\prd] {10.1103/PhysRevD.104.023524}, \href
  {https://ui.adsabs.harvard.edu/abs/2021PhRvD.104b3524H} {104, 023524}

\bibitem[\protect\citeauthoryear{{Hindmarsh}, {Huber}, {Rummukainen}  \&
  {Weir}}{{Hindmarsh} et~al.}{2015}]{2015PhRvD..92l3009H}
{Hindmarsh} M.,  {Huber} S.~J.,  {Rummukainen} K.,   {Weir} D.~J.,  2015,
  \mn@doi [\prd] {10.1103/PhysRevD.92.123009}, \href
  {https://ui.adsabs.harvard.edu/abs/2015PhRvD..92l3009H} {92, 123009}

\bibitem[\protect\citeauthoryear{{Huang} et~al.,}{{Huang}
  et~al.}{2020}]{2020PhRvD.102f3021H}
{Huang} S.-J.,  et~al., 2020, \mn@doi [\prd] {10.1103/PhysRevD.102.063021},
  \href {https://ui.adsabs.harvard.edu/abs/2020PhRvD.102f3021H} {102, 063021}

\bibitem[\protect\citeauthoryear{{Kalita} \& {Mukhopadhyay}}{{Kalita} \&
  {Mukhopadhyay}}{2019}]{2019MNRAS.490.2692K}
{Kalita} S.,  {Mukhopadhyay} B.,  2019, \mn@doi [\mnras]
  {10.1093/mnras/stz2734}, \href
  {https://ui.adsabs.harvard.edu/abs/2019MNRAS.490.2692K} {490, 2692}

\bibitem[\protect\citeauthoryear{{Kalita}, {Mukhopadhyay}, {Mondal}  \&
  {Bulik}}{{Kalita} et~al.}{2020}]{2020ApJ...896...69K}
{Kalita} S.,  {Mukhopadhyay} B.,  {Mondal} T.,   {Bulik} T.,  2020, \mn@doi
  [\apj] {10.3847/1538-4357/ab8e40}, \href
  {https://ui.adsabs.harvard.edu/abs/2020ApJ...896...69K} {896, 69}

\bibitem[\protect\citeauthoryear{{Kalita}, {Mondal}, {Tout}, {Bulik}  \&
  {Mukhopadhyay}}{{Kalita} et~al.}{2021}]{2021MNRAS.508..842K}
{Kalita} S.,  {Mondal} T.,  {Tout} C.~A.,  {Bulik} T.,   {Mukhopadhyay} B.,
  2021, \mn@doi [\mnras] {10.1093/mnras/stab2625}, \href
  {https://ui.adsabs.harvard.edu/abs/2021MNRAS.508..842K} {508, 842}

\bibitem[\protect\citeauthoryear{{Kashiyama}, {Ioka}  \&
  {M{\'e}sz{\'a}ros}}{{Kashiyama} et~al.}{2013}]{2013ApJ...776L..39K}
{Kashiyama} K.,  {Ioka} K.,   {M{\'e}sz{\'a}ros} P.,  2013, \mn@doi [\apjl]
  {10.1088/2041-8205/776/2/L39}, \href
  {https://ui.adsabs.harvard.edu/abs/2013ApJ...776L..39K} {776, L39}

\bibitem[\protect\citeauthoryear{{Kolosnitsyn} \& {Rudenko}}{{Kolosnitsyn} \&
  {Rudenko}}{2015}]{2015PhyS...90g4059K}
{Kolosnitsyn} N.~I.,  {Rudenko} V.~N.,  2015, \mn@doi [\physscr]
  {10.1088/0031-8949/90/7/074059}, \href
  {https://ui.adsabs.harvard.edu/abs/2015PhyS...90g4059K} {90, 074059}

\bibitem[\protect\citeauthoryear{{Komatsu}, {Eriguchi}  \& {Hachisu}}{{Komatsu}
  et~al.}{1989}]{1989MNRAS.237..355K}
{Komatsu} H.,  {Eriguchi} Y.,   {Hachisu} I.,  1989, \mn@doi [\mnras]
  {10.1093/mnras/237.2.355}, \href
  {http://adsabs.harvard.edu/abs/1989MNRAS.237..355K} {237, 355}

\bibitem[\protect\citeauthoryear{{Kulkarni}, {Ofek}, {Neill}, {Zheng}  \&
  {Juric}}{{Kulkarni} et~al.}{2014}]{2014ApJ...797...70K}
{Kulkarni} S.~R.,  {Ofek} E.~O.,  {Neill} J.~D.,  {Zheng} Z.,   {Juric} M.,
  2014, \mn@doi [\apj] {10.1088/0004-637X/797/1/70}, \href
  {https://ui.adsabs.harvard.edu/abs/2014ApJ...797...70K} {797, 70}

\bibitem[\protect\citeauthoryear{{Kumar}, {Lu}  \& {Bhattacharya}}{{Kumar}
  et~al.}{2017}]{2017MNRAS.468.2726K}
{Kumar} P.,  {Lu} W.,   {Bhattacharya} M.,  2017, \mn@doi [\mnras]
  {10.1093/mnras/stx665}, \href
  {https://ui.adsabs.harvard.edu/abs/2017MNRAS.468.2726K} {468, 2726}

\bibitem[\protect\citeauthoryear{{Kuroda}, {Ni}  \& {Pan}}{{Kuroda}
  et~al.}{2015}]{2015IJMPD..2430031K}
{Kuroda} K.,  {Ni} W.-T.,   {Pan} W.-P.,  2015, \mn@doi [International Journal
  of Modern Physics D] {10.1142/S0218271815300311}, \href
  {https://ui.adsabs.harvard.edu/abs/2015IJMPD..2430031K} {24, 1530031}

\bibitem[\protect\citeauthoryear{{Kushwaha}, {Malik}  \&
  {Shankaranarayanan}}{{Kushwaha} et~al.}{2022}]{2022arXiv220200032K}
{Kushwaha} A.,  {Malik} S.,   {Shankaranarayanan} S.,  2022, arXiv e-prints,
  \href {https://ui.adsabs.harvard.edu/abs/2022arXiv220200032K} {p.
  arXiv:2202.00032}

\bibitem[\protect\citeauthoryear{{Leaci}, {LIGO Scientific Collaboration}  \&
  {Virgo Collaboration}}{{Leaci} et~al.}{2012}]{2012JPhCS.354a2010L}
{Leaci} P.,  {LIGO Scientific Collaboration}  {Virgo Collaboration} 2012,
  \mn@doi [Journal of Physics: Conference Series]
  {10.1088/1742-6596/354/1/012010}, \href
  {https://ui.adsabs.harvard.edu/abs/2012JPhCS.354a2010L} {354, 012010}

\bibitem[\protect\citeauthoryear{{Liu}}{{Liu}}{2020}]{2020IJAA...10...28L}
{Liu} X.,  2020, \mn@doi [International Journal of Astronomy and Astrophysics]
  {10.4236/ijaa.2020.101003}, \href
  {https://ui.adsabs.harvard.edu/abs/2020IJAA...10...28L} {10, 28}

\bibitem[\protect\citeauthoryear{{Lorimer}, {Bailes}, {McLaughlin}, {Narkevic}
  \& {Crawford}}{{Lorimer} et~al.}{2007}]{2007Sci...318..777L}
{Lorimer} D.~R.,  {Bailes} M.,  {McLaughlin} M.~A.,  {Narkevic} D.~J.,
  {Crawford} F.,  2007, \mn@doi [Science] {10.1126/science.1147532}, \href
  {https://ui.adsabs.harvard.edu/abs/2007Sci...318..777L} {318, 777}

\bibitem[\protect\citeauthoryear{{Lu} \& {Kumar}}{{Lu} \&
  {Kumar}}{2019}]{2019MNRAS.483L..93L}
{Lu} W.,  {Kumar} P.,  2019, \mn@doi [\mnras] {10.1093/mnrasl/sly200}, \href
  {https://ui.adsabs.harvard.edu/abs/2019MNRAS.483L..93L} {483, L93}

\bibitem[\protect\citeauthoryear{{Lyubarsky}}{{Lyubarsky}}{2014}]{2014MNRAS.442L...9L}
{Lyubarsky} Y.,  2014, \mn@doi [\mnras] {10.1093/mnrasl/slu046}, \href
  {https://ui.adsabs.harvard.edu/abs/2014MNRAS.442L...9L} {442, L9}

\bibitem[\protect\citeauthoryear{Maggiore}{Maggiore}{2008}]{Maggiore}
Maggiore M.,  2008, Gravitational waves: Volume 1: Theory and experiments.
Oxford Master Series in Physics, Oxford university press,
  \mn@doi{10.1093/acprof:oso/9780198570745.001.0001}

\bibitem[\protect\citeauthoryear{{Melatos}}{{Melatos}}{2000}]{2000MNRAS.313..217M}
{Melatos} A.,  2000, \mn@doi [\mnras] {10.1046/j.1365-8711.2000.03031.x}, \href
  {https://ui.adsabs.harvard.edu/abs/2000MNRAS.313..217M} {313, 217}

\bibitem[\protect\citeauthoryear{{Miller} \& {Yunes}}{{Miller} \&
  {Yunes}}{2019}]{2019Natur.568..469M}
{Miller} M.~C.,  {Yunes} N.,  2019, \mn@doi [\nat] {10.1038/s41586-019-1129-z},
  \href {https://ui.adsabs.harvard.edu/abs/2019Natur.568..469M} {568, 469}

\bibitem[\protect\citeauthoryear{{Moore}, {Cole}  \& {Berry}}{{Moore}
  et~al.}{2015}]{2015CQGra..32a5014M}
{Moore} C.~J.,  {Cole} R.~H.,   {Berry} C.~P.~L.,  2015, \mn@doi [Classical and
  Quantum Gravity] {10.1088/0264-9381/32/1/015014}, \href
  {https://ui.adsabs.harvard.edu/abs/2015CQGra..32a5014M} {32, 015014}

\bibitem[\protect\citeauthoryear{{Patel}, {Siemens}, {Dupuis}  \&
  {Betzwieser}}{{Patel} et~al.}{2010}]{2010PhRvD..81h4032P}
{Patel} P.,  {Siemens} X.,  {Dupuis} R.,   {Betzwieser} J.,  2010, \mn@doi
  [\prd] {10.1103/PhysRevD.81.084032}, \href
  {https://ui.adsabs.harvard.edu/abs/2010PhRvD..81h4032P} {81, 084032}

\bibitem[\protect\citeauthoryear{{Petroff} et~al.,}{{Petroff}
  et~al.}{2016}]{2016PASA...33...45P}
{Petroff} E.,  et~al., 2016, \mn@doi [\pasa] {10.1017/pasa.2016.35}, \href
  {https://ui.adsabs.harvard.edu/abs/2016PASA...33...45P} {33, e045}

\bibitem[\protect\citeauthoryear{{Philippov}, {Spitkovsky}  \&
  {Cerutti}}{{Philippov} et~al.}{2015}]{2015ApJ...801L..19P}
{Philippov} A.~A.,  {Spitkovsky} A.,   {Cerutti} B.,  2015, \mn@doi [\apjl]
  {10.1088/2041-8205/801/1/L19}, \href
  {https://ui.adsabs.harvard.edu/abs/2015ApJ...801L..19P} {801, L19}

\bibitem[\protect\citeauthoryear{{Pili}, {Bucciantini}  \& {Del Zanna}}{{Pili}
  et~al.}{2014}]{2014MNRAS.439.3541P}
{Pili} A.~G.,  {Bucciantini} N.,   {Del Zanna} L.,  2014, \mn@doi [\mnras]
  {10.1093/mnras/stu215}, \href
  {http://adsabs.harvard.edu/abs/2014MNRAS.439.3541P} {439, 3541}

\bibitem[\protect\citeauthoryear{{Pili}, {Bucciantini}  \& {Del Zanna}}{{Pili}
  et~al.}{2017}]{2017MNRAS.470.2469P}
{Pili} A.~G.,  {Bucciantini} N.,   {Del Zanna} L.,  2017, \mn@doi [\mnras]
  {10.1093/mnras/stx1176}, \href
  {https://ui.adsabs.harvard.edu/abs/2017MNRAS.470.2469P} {470, 2469}

\bibitem[\protect\citeauthoryear{{Pitkin}}{{Pitkin}}{2011}]{2011MNRAS.415.1849P}
{Pitkin} M.,  2011, \mn@doi [\mnras] {10.1111/j.1365-2966.2011.18818.x}, \href
  {https://ui.adsabs.harvard.edu/abs/2011MNRAS.415.1849P} {415, 1849}

\bibitem[\protect\citeauthoryear{{Platts}, {Weltman}, {Walters}, {Tendulkar},
  {Gordin}  \& {Kandhai}}{{Platts} et~al.}{2019}]{2019PhR...821....1P}
{Platts} E.,  {Weltman} A.,  {Walters} A.,  {Tendulkar} S.~P.,  {Gordin}
  J.~E.~B.,   {Kandhai} S.,  2019, \mn@doi [\physrep]
  {10.1016/j.physrep.2019.06.003}, \href
  {https://ui.adsabs.harvard.edu/abs/2019PhR...821....1P} {821, 1}

\bibitem[\protect\citeauthoryear{{Pons}, {Vigan{\`o}}  \& {Geppert}}{{Pons}
  et~al.}{2012}]{2012A&A...547A...9P}
{Pons} J.~A.,  {Vigan{\`o}} D.,   {Geppert} U.,  2012, \mn@doi [\aap]
  {10.1051/0004-6361/201220091}, \href
  {https://ui.adsabs.harvard.edu/abs/2012A&A...547A...9P} {547, A9}

\bibitem[\protect\citeauthoryear{{Portilla} \& {Lapiedra}}{{Portilla} \&
  {Lapiedra}}{2001}]{2001PhRvD..63d4014P}
{Portilla} M.,  {Lapiedra} R.,  2001, \mn@doi [\prd]
  {10.1103/PhysRevD.63.044014}, \href
  {https://ui.adsabs.harvard.edu/abs/2001PhRvD..63d4014P} {63, 044014}

\bibitem[\protect\citeauthoryear{{Pustovoit}, {Gladyshev}, {Kauts}, {Morozov},
  {Nikolaev}, {Fomin}, {Sharandin}  \& {Kayutenko}}{{Pustovoit}
  et~al.}{2021}]{2021JPhCS2081a2009P}
{Pustovoit} V.,  {Gladyshev} V.,  {Kauts} V.,  {Morozov} A.,  {Nikolaev} P.,
  {Fomin} I.,  {Sharandin} E.,   {Kayutenko} A.,  2021, in Journal of Physics
  Conference Series. p. 012009, \mn@doi{10.1088/1742-6596/2081/1/012009}

\bibitem[\protect\citeauthoryear{{Servin} \& {Brodin}}{{Servin} \&
  {Brodin}}{2003}]{2003PhRvD..68d4017S}
{Servin} M.,  {Brodin} G.,  2003, \mn@doi [\prd] {10.1103/PhysRevD.68.044017},
  \href {https://ui.adsabs.harvard.edu/abs/2003PhRvD..68d4017S} {68, 044017}

\bibitem[\protect\citeauthoryear{{Spitkovsky}}{{Spitkovsky}}{2006}]{2006ApJ...648L..51S}
{Spitkovsky} A.,  2006, \mn@doi [\apjl] {10.1086/507518}, \href
  {https://ui.adsabs.harvard.edu/abs/2006ApJ...648L..51S} {648, L51}

\bibitem[\protect\citeauthoryear{{Stephenson}}{{Stephenson}}{2005}]{2005AIPC..746.1264S}
{Stephenson} G.~V.,  2005, in {El-Genk} M.~S.,  ed.,  American Institute of
  Physics Conference Series Vol. 746, Space Technology and Applications
  International Forum - STAIF 2005. pp 1264--1270, \mn@doi{10.1063/1.1867254}

\bibitem[\protect\citeauthoryear{{Sun} \& {Zhang}}{{Sun} \&
  {Zhang}}{2021}]{2021PhRvD.104j3009S}
{Sun} S.,  {Zhang} Y.-L.,  2021, \mn@doi [\prd] {10.1103/PhysRevD.104.103009},
  \href {https://ui.adsabs.harvard.edu/abs/2021PhRvD.104j3009S} {104, 103009}

\bibitem[\protect\citeauthoryear{{Totani}}{{Totani}}{2013}]{2013PASJ...65L..12T}
{Totani} T.,  2013, \mn@doi [\pasj] {10.1093/pasj/65.5.L12}, \href
  {https://ui.adsabs.harvard.edu/abs/2013PASJ...65L..12T} {65, L12}

\bibitem[\protect\citeauthoryear{{Wang}, {Luo}, {Yue}, {Chen}, {Lee}  \&
  {Xu}}{{Wang} et~al.}{2018}]{2018ApJ...852..140W}
{Wang} W.,  {Luo} R.,  {Yue} H.,  {Chen} X.,  {Lee} K.,   {Xu} R.,  2018,
  \mn@doi [\apj] {10.3847/1538-4357/aaa025}, \href
  {https://ui.adsabs.harvard.edu/abs/2018ApJ...852..140W} {852, 140}

\bibitem[\protect\citeauthoryear{{Wickramasinghe}, {Tout}  \&
  {Ferrario}}{{Wickramasinghe} et~al.}{2014}]{2014MNRAS.437..675W}
{Wickramasinghe} D.~T.,  {Tout} C.~A.,   {Ferrario} L.,  2014, \mn@doi [\mnras]
  {10.1093/mnras/stt1910}, \href
  {https://ui.adsabs.harvard.edu/abs/2014MNRAS.437..675W} {437, 675}

\bibitem[\protect\citeauthoryear{{Zel'dovich}}{{Zel'dovich}}{1974}]{1974JETP...38..652Z}
{Zel'dovich} Y.~B.,  1974, Soviet Journal of Experimental and Theoretical
  Physics, \href {https://ui.adsabs.harvard.edu/abs/1974JETP...38..652Z} {38,
  652}

\bibitem[\protect\citeauthoryear{{Zheng}, {Wei}, {Wen}  \& {Li}}{{Zheng}
  et~al.}{2018}]{2018PhRvD..98f4028Z}
{Zheng} H.,  {Wei} L.~F.,  {Wen} H.,   {Li} F.~Y.,  2018, \mn@doi [\prd]
  {10.1103/PhysRevD.98.064028}, \href
  {https://ui.adsabs.harvard.edu/abs/2018PhRvD..98f4028Z} {98, 064028}

\bibitem[\protect\citeauthoryear{{Zimmermann} \& {Szedenits}}{{Zimmermann} \&
  {Szedenits}}{1979}]{1979PhRvD..20..351Z}
{Zimmermann} M.,  {Szedenits} E. J.,  1979, \mn@doi [\prd]
  {10.1103/PhysRevD.20.351}, \href
  {https://ui.adsabs.harvard.edu/abs/1979PhRvD..20..351Z} {20, 351}

\makeatother
\end{thebibliography}

\bsp	
\label{lastpage}
\end{document}